\documentclass[prb,aps,tightenlines,twocolumn,
showpacs,floatfix]{revtex4}
\usepackage{graphicx}
\usepackage{amssymb}
\usepackage{amsmath}
\usepackage{bm}
\usepackage{colordvi}

\begin{document}
\title{Electron spin relaxation in graphene from a  microscopic approach:
 Role of electron-electron interaction}
\author{Y. Zhou}
\author{M. W. Wu}
\thanks{Author to whom correspondence should be addressed}
\email{mwwu@ustc.edu.cn.}
\affiliation{Hefei National Laboratory for Physical Sciences at
  Microscale and Department of Physics, University of Science and
  Technology of China, Hefei, Anhui, 230026, China}

\date{\today}
\begin{abstract}
Electron spin relaxation in graphene on a substrate is
investigated from the  microscopic kinetic spin Bloch equation
approach. All the relevant scatterings, such as the electron-impurity,
electron--acoustic-phonon, electron--optical-phonon,
electron--remote-interfacial-phonon, as
well as electron-electron Coulomb scatterings, are explicitly included.
Our study concentrates on clean intrinsic graphene, where the
spin-orbit coupling from the adatoms can be neglected.
We discuss the effect of the electron-electron Coulomb interaction on
spin relaxation under various conditions.
It is shown that the electron-electron Coulomb scattering plays an
important role in spin relaxation at high temperature. 
We also find a significant increase of the spin relaxation time
for high spin polarization even at room temperature,
which is due to the Coulomb 
Hartree-Fock contribution-induced  effective longitudinal magnetic field.
It is also discovered that the spin relaxation time increases with the
in-plane electric field due to the hot-electron effect, which is
different from the non-monotonic behavior in semiconductors.
Moreover, we show that  the electron-electron Coulomb scattering
in graphene is not strong enough to establish the steady-state hot-electron
distribution widely used in the literature and an 
alternative approximate one is proposed based on our computation.
\end{abstract}


\maketitle

\section{Introduction}
Graphene, as a strictly two-dimensional  material, has revealed a
cornucopia of new physics and potential applications, and thus has
attracted much attention in recent years.\cite{Geim_07}
This material is also important for spintronics since the spin
relaxation time (SRT) in intrinsic graphene is expected to be very
long. The underlying reason is the low hyperfine interaction of
the spin with the carbon nuclei (natural carbon only 
contains 1\% $^{13}$C with spin) and the weak spin-orbit coupling (SOC) 
due to the low atomic number.\cite{Loss_hyper,Kane_SOC,
Hernando_06,Min,Fang_07,Fabian_SOC}

The study of spin relaxation in graphene is still in the initial stage. Some
investigations have been performed on the spin
relaxation due to the D'yakonov-Perel' (DP) mechanism\cite{Dyakonov} 
in graphene.\cite{Fabian_SR,Hernando_07,Hernando_09}  
This relaxation mechanism is from the joint
effects of the momentum scattering and the momentum-dependent
spin-orbit field (inhomogeneous broadening\cite{Wu_01,Wu_rev}).
However, the previous investigations are all in the framework of single-particle
approach, thus the electron-electron Coulomb scattering, which
has been demonstrated to be very important for the spin relaxation in bulk and
low-dimensional semiconductor systems,\cite{Wu_01,Wu_rev,Zhou_PRB_07,
Jiang_bulk_09,highP,wu-exp-hP,tobias,Zheng_exp,hot-e,multi,Ivchenko,Leyland,Ji} 
was not included. Understanding the effect of the
electron-electron Coulomb scattering on spin relaxation in graphene is an
important problem. 
In addition, the Coulomb Hartree-Fock (HF) term acts as an 
effective longitudinal
magnetic field, and hence can increase the SRT by more than one order of
magnitude for high initial spin polarization in semiconductors at low 
temperature.\cite{highP,wu-exp-hP,tobias,Zheng_exp,Jiang_bulk_09}
Whether it is still valid in graphene remains unchecked.
Also, in semiconductors the spin relaxation can be effectively
manipulated by the high in-plane electric 
field.\cite{hot-e,multi,Jiang_bulk_09} 
How the in-plane electric field affects the spin relaxation in graphene is also
unclear.
In the present paper, we investigate the spin relaxation in graphene from the 
microscopic kinetic spin Bloch equation (KSBE) approach,\cite{Wu_rev}
which has achieved much success in the study of the spin dynamics in 
semiconductors.
Via this approach, we can explicitly include all the relevant scatterings,
especially the electron-electron Coulomb scattering, 
and understand the physics raised above. 

It is also noted that there is a significant discrepancy between the
existing theories and the recent spin transport
experiments.\cite{Tombros_07,Tombros_08,Popinciuc,Jozsa_09} 
These experiments reported the SRTs of only about $150$~ps, at least
one order of magnitude shorter than the lowest value obtained in
the theory.\cite{Fabian_SR,Hernando_07,Hernando_09} 
This suggests that the SRTs obtained in the recent
experiments are likely to be limited by an extrinsic
mechanisms, e.g., the local spin-orbit field from the 
adatoms.\cite{Fabian_SR,Hernando_09,Castro_imp}
In this paper we concentrate on the relatively cleaner graphene
samples by choosing low impurity densities which give a mobility
higher than the values in the latest experiments, 
so that the effect of the adatoms can be neglected.

This paper is organized as follows. In Sec.~II, we present the model
and introduce the KSBEs. Then in Sec.~IIIA, we discuss the
effect of the electron-electron Coulomb interaction on spin relaxation
at various temperature, electron density, initial spin polarization and
in-plane electric field. We discuss the hot-electron
distribution function in the steady state in Sec.~IIIB. 
Finally, we summarize  in Sec.~IV.

\section{Model and KSBEs}
We start our investigation from a graphene layer on a SiO$_2$ substrate. The  
$z$-axis is set perpendicular to the graphene plane. A
uniform electric field ${\bf E}_{\|}$ and a uniform magnetic field
${\bf B}$ are applied along the $x$- and $y$-axes respectively (the
Voigt configuration). 
Without the SOC and the external field, the band
structure of graphene near the $K$ and $K^\prime$ points can be
described by the effective Hamiltonian ($\hbar\equiv 1$)\cite{DiVincenzo} 
\begin{equation}
  H_0^\mu=v_{\rm F}(\mu\sigma_x k_x+ \sigma_y k_y).
  \label{H_0}
\end{equation}
Here $\mu=1(-1)$ for $K(K^\prime)$ valleys;
$v_{\rm F}$ is the Fermi velocity; ${\bf k}$ represents the two-dimensional
wave vector relative to $K(K^\prime)$ points; 
$\bm{\sigma}$ is the Pauli matrix in the pseudospin
space formed by the A and B sublattices of the honeycomb lattice. 
The eigenvalues of $H_0^\mu$ are
$\epsilon_{\mu\nu{\bf k}}=\nu v_{\rm F}|{\bf k}|$ with $\nu=1(-1)$ for
electron (hole) band. The corresponding eigenstates are 
$\psi_{{\bf k}}^{\mu\nu}=1/\sqrt{2}(\mu\nu
e^{-i\mu\theta_{\bf k}},1)^{\rm T}$ with $\theta_{\bf k}$
representing the polar angle of ${\bf k}$.
We introduce an orthogonal and complete basis set 
$\Psi_{{\bf k}s}^{\mu\nu}=\psi_{{\bf k}}^{\mu\nu}
\otimes\chi_{s}$ with $\chi_{s}$
being the eigenstate of the spin Pauli matrix $s_z$. 
In this basis set, the total Hamiltonian including the 
SOC can be written as\cite{Fabian_SR,correct} 
\begin{eqnarray}
  \nonumber
  {H}_\mathrm{eff} &=& \sum_{{\mu}\nu{\bf k}ss^\prime}
  \big[(\epsilon_{\mu\nu{\bf k}}-\lambda_{\rm I}
  -e{\bf E}_{\|}\cdot {\bf R})
  \delta_{ss^\prime}\\ &&
  +(g\mu_B{\bf B} + \nu \bm{\Omega}_{\bf k})
  \cdot\mathbf{s}_{ss^\prime}\big] {c^{{\mu}\nu}_{{\bf k}s}}^{\dagger}
  c^{{\mu}\nu}_{{\bf k}s^\prime} + {H}_{\rm int}.
  \label{H_eff}
\end{eqnarray}
where ${\bf R}=(x,y)$ is the electron position.
$c^{{\mu}\nu}_{{\bf k}s}$ (${c^{{\mu}\nu}_{{\bf k}s}}^{\dagger}$) is the
annihilation (creation) operator of the state $\Psi_{{\bf k}s}^{\mu\nu}$.
$e$ is the electron charge ($e>0$). $g=2$ is the effective Land\'e
factor. The intrinsic SOC coefficient $\lambda_{\rm I}=0.012$~meV is 
from the recent first-principle calculation.\cite{Fabian_SOC} From 
Eq.~(\ref{H_eff}), it is seen that the intrinsic SOC only induces a
shift of the energy spectrum of graphene.
$\bm{\Omega}_{\bf k}$ denotes the effective magnetic field due
to the Rashba SOC, which reads\cite{Fabian_SR}
\begin{equation}
  \bm{\Omega}_{\bf  k}=\alpha_{\rm R}(-\sin\theta_{\bf k},
  \cos\theta_{\bf k},0)
\end{equation}
with $\alpha_{\rm R}=\zeta E_z$. The recent first-principle
calculation gives $\zeta=0.005$~meV$\cdot$nm/V.\cite{Fabian_SOC}
The longitudinal electric field $E_z$ originates
from the gate voltage and chemical doping. 
Here we choose a typical value in experiment
$E_z=300$~kV/cm.\cite{Novoselov_Science,Ez_sub} 
It is noted that $\bm{\Omega}_{\bf k}$  depends on the
direction of ${\bf k}$ only, 
but is independent on the magnitude of ${\bf k}$. Therefore the inhomogeneous
broadening induced by the SOC does not change with the variation of
temperature and electron density. This makes the temperature and 
electron-density dependences of the SRT in graphene  very different
from those in semiconductors.\cite{Wu_rev} 
The interaction Hamiltonian $H_{\rm int}$ consists of the electron-impurity,
electron-phonon as well as electron-electron Coulomb interactions. Their
expressions are given in the appendix. 
In the derivation of Eq.~(\ref{H_eff}),
$|\epsilon_{\mu\nu{\bf k}}|\gg \alpha_{\rm R}+\lambda_{\rm I}$ is assumed and
thus the terms between the electron and hole bands are neglected. This
approximation is valid when the Fermi energy $E_{\rm F}$ is much
larger than $0.03$~meV,\cite{Fabian_SOC,Fabian_SR} which is usually
fulfilled in gated or doped graphene. 
In our calculation, we restrict ourselves to the $n$-doped case
(i.e., $E_{\rm F}\gg k_{\rm B}T$). 

By using the nonequilibrium Green function method,\cite{Haug_1998}
the KSBEs can be constructed as:\cite{Wu_rev}
\begin{equation}
  \partial_t\hat{\rho}_{{\mu}{\bf k}}= \left.\partial_t\hat{\rho}_{{\mu}{\bf k}}\right|_
  {\rm coh}+\left.\partial_t\hat{\rho}_{{\mu}{\bf k}}\right|_{\rm drift}
  +\left.\partial_t\hat{\rho}_{{\mu}{\bf k}}\right|_{\rm scat},
\end{equation}
where $\hat{\rho}_{{\mu}\bf k}$ represent the density matrices of
electron with the relative momentum ${\bf k}$ in valley ${\mu}$, 
whose diagonal terms $\rho_{\mu{\bf k},ss}\equiv f_{{\mu}{\bf k}s}$
($s=\pm\frac{1}{2}$) represent the electron distribution functions
and off-diagonal ones $\rho_{\mu{\bf k},\frac{1}{2}\,{-}\frac{1}{2}}=
\rho_{\mu{\bf k},{-}\frac{1}{2}\,\frac{1}{2}}^\ast$ describe the spin coherence.
The coherent term is given by
\begin{equation}
\left.\partial_t\hat{\rho}_{{\mu}{\bf k}}\right|_{\rm coh}=
  -i\left[(g\mu_B{\bf B}+\bm{\Omega}_{\bf k})\cdot\hat{\bf s}
    +\hat{\Sigma}^{\rm HF}_{\mu{\bf k}},\;\; \hat{\rho}_{{\mu}{\bf k}} \right],
  \label{coherent}
\end{equation}
in which $[A,B]\equiv AB-BA$ is the commutator; $\hat{\Sigma}^\mathrm{HF}
_{\mu{\bf k}}=-\sum_{{\bf k}^\prime}V^{\mu 11}_{{\bf k},{\bf k}^\prime}
I_{{\bf k}{\bf k}^\prime}\hat{\rho}_{{\mu}{\bf k}^\prime}$
is the effective magnetic field from the Coulomb HF
contribution.\cite{highP}
The drift term can be written as\cite{hot-e}
\begin{equation}
\left.\partial_{t}\hat{\rho}_{{\mu}{\bf k}}\right|_{\rm drift} = e 
  {\bf E}_{\|}\cdot\mbox{\boldmath$\nabla$\unboldmath}_{\bf k} 
  \hat{\rho}_{\mu{\bf k}},
\end{equation}
The scattering term $\partial_t\hat{\rho}_{{\mu}{\bf k}}|_{\rm scat}$
consists of the electron-impurity, electron--acoustic (AC)-phonon,
electron--optical (OP)-phonon, 
electron--remote-interfacial (RI)-phonon as well as electron-electron
Coulomb scatterings. These scattering terms are
\begin{widetext}
\begin{eqnarray}
  \left. \partial_{t}\hat{\rho}_{{\mu}{\bf k}}\right|_{\rm ei} &=&
  - \pi N_i \sum_{{\bf k}^{\prime}} |U^{\mu 1}_{{\bf k},{\bf k}^{\prime}}|^2
  I_{{\bf k}{\bf k}^\prime}  
  \delta(\varepsilon_{{\mu}{\bf k}^{\prime}}-\varepsilon_{{\mu}{\bf  k}})  
  \left( \hat{\rho}^{>}_{\mu{\bf k}^{\prime}} \hat{\rho}^{<}
    _{\mu{\bf k}} - \hat{\rho}^{<}_{\mu{\bf k}^{\prime}}
    \hat{\rho}^{>}_{\mu{\bf k}} \right)  + {\rm H.c.},
  \label{scat_ei}\\
  \left. \partial_{t}\hat{\rho}_{{\mu}{\bf k}}\right|_{\rm ep} &=&
  - \pi \sum_{\mu^\prime{\bf k}^{\prime}\lambda,\pm}
  |M^{\lambda}_{\mu{\bf k},\mu^\prime{\bf k}^{\prime}}|^2
  \delta(\varepsilon_{{\mu^\prime}{\bf k}^{\prime}}
  -\varepsilon_{{\mu}{\bf k}}\pm \omega^{\lambda}_{{\bf k}-{\bf k}^{\prime}})
   \left( N^{\pm}_{\lambda,{\bf k}-{\bf k}^{\prime}}
    \hat{\rho}^{>}_{\mu^\prime{\bf k}^{\prime}} \hat{\rho}^{<}_{\mu{\bf k}} -
    N^{\mp}_{\lambda,{\bf k}-{\bf k}^{\prime}}
    \hat{\rho}^{<}_{\mu^\prime{\bf k}^{\prime}} \hat{\rho}^{>}
    _{\mu{\bf k}} \right)  + {\rm H.c.},
  \label{scat_ep}\\
  \nonumber
  \left. \partial_{t}\hat{\rho}_{{\mu}{\bf k}}\right|_{\rm ee} &=& 
  - \pi \sum_{\mu^\prime{\bf k}^{\prime}{\bf k}^{\prime\prime}}
  |V^{\mu 11}_{{\bf k},{\bf k}^{\prime}}|^2
  I_{{\bf k}{\bf k}^\prime}  I_{{\bf k}^{\prime\prime}
    \mathtt{{\bf k}^{\prime\prime}{-}{\bf k}{+}{\bf k}^{\prime}}}
  \delta(\varepsilon_{{\mu}{\bf k}^{\prime}}{-}\varepsilon_{{\mu}{\bf
      k}}{+}\varepsilon_{{\mu}^\prime{\bf k}^{\prime\prime}}{-}\varepsilon
    _{\mathtt{\mu^\prime{\bf k}^{\prime\prime}-{\bf k}+{\bf k}^{\prime}}})
    \Big[ {\rm Tr}\left(
    \hat{\rho}^{<}_{\mu^\prime{\bf k}^{\prime\prime}{-}{\bf k}{+}{\bf k}^{\prime}}
    \hat{\rho}^{>}_{\mu^\prime{\bf k}^{\prime\prime}}\right)
  \hat{\rho}^{>}_{\mu{\bf k}^{\prime}}\hat{\rho}^{<}_{\mu{\bf k}}
  \\ && 
  \mbox{} - {\rm Tr}\left(
    \hat{\rho}^{>}_{\mu^\prime{\bf k}^{\prime\prime}{-}{\bf k}{+}{\bf k}^{\prime}}
    \hat{\rho}^{<}_{\mu^\prime{\bf k}^{\prime\prime}}\right)
  \hat{\rho}^{<}_{\mu{\bf k}^{\prime}}
  \hat{\rho}^{>}_{\mu{\bf k}}  \Big]
  + {\rm H.c.}.
  \label{scat_ee}
\end{eqnarray}
\end{widetext}
In these equations, $\varepsilon_{{\mu}{\bf k}}\equiv
\epsilon_{\mu\nu=1{\bf k}}=v_{\rm F}|{\bf k}|$,\cite{linear}
$\hat{\rho}^>_{{\mu}{\bf k}}\equiv1-\hat{\rho}_{{\mu}{\bf k}}$,  
$\hat{\rho}^<_{{\mu}{\bf k}}\equiv\hat{\rho}_{{\mu}{\bf k}}$;
$\omega^{\lambda}_{\bf q}$ denotes the phonon energy spectrum;
$N_{\lambda{\bf q}}^{\pm}=N_{\lambda{\bf q}}
+\frac{1}{2}\pm\frac{1}{2}$ with $N_{\lambda{\bf q}}$ representing the
phonon number at lattice temperature.
The form factor $I_{{\bf k}{\bf k}-{\bf q}}=
|{\psi_{{\bf k}}^{\mu1}}^\dagger\psi_{{\bf k}-{\bf q}}^{\mu1}|^2
=\frac{1}{2}
[1+\cos(\theta_{\bf k}-\theta_{{\bf k}-{\bf q}})]$.
The matrix elements $U^{\mu 1}_{{\bf k},{\bf k}^{\prime}}$, 
$V^{\mu 11}_{{\bf k},{\bf k}^{\prime}}$ and $M^{\lambda}_{\mu{\bf k},
\mu^\prime{\bf k}^{\prime}}$ are given in the appendix.

\section{Results}
The KSBEs with all the scatterings explicitly included can be solved
self-consistently following the numerical scheme similar to that
in semiconductors, detailed  in the appendix of
Ref.~\onlinecite{hot-e}. Then one can obtain the temporal evolution of
the electron density matrix. The SRT $T_1$ and ensemble spin dephasing
time $T_2^\ast$ can be determined from the slopes of the
envelopes of $\Delta N(t)=\sum_{\mu{\bf k}}(f_{\mu{\bf k}\frac{1}{2}}
-f_{\mu{\bf k}-\frac{1}{2}})$ and $\rho(t)=\left|\sum_{\mu{\bf k}}
\rho_{\mu{\bf k}\frac{1}{2}\,-\frac{1}{2}}\right|$,
respectively.\cite{Lv_PLA,Wu_rev} 
Since the Rashba spin-orbit field is 
in the graphene plane, these two SRTs satisfy
$T_2^\ast=2T_1$ in the case without magnetic field. 
In the presence of an in-plane magnetic field, the spin
relaxation becomes isotropic, i.e.,
 $T_1=T_2^\ast=2/(T_1({\bf B}=0)^{-1}+T_2^\ast({\bf B}=0)^{-1})
=\frac{4}{3}T_1({\bf B}=0)$.\cite{Wu_rev} In the following, we only
show the SRT $\tau\equiv T_1$. Unless otherwise specified, we choose
initial spin polarization $P=1$~\%, electron density $N_e=7\times
10^{11}$~cm$^{-2}$ ($E_{\rm F}=100$~meV), external magnetic field
${\bf B}=0$ and electric field ${\bf E}_{\|}=0$.
The effective impurity density is chosen to be $N_i=2\times 10^{11}$~cm$^{-2}$.
The corresponding mobility is $\mu=3\times 10^{4}$~cm$^2$/V$\cdot$s at $100$~K,
which is of the same order of magnitude as those reported in the
experiment\cite{Novoselov_Science} but one order of magnitude higher
than those in the recent spin transport
experiments.\cite{Tombros_07,Tombros_08,Popinciuc,Jozsa_09} 

\subsection{Effect of electron-electron Coulomb interaction on spin relaxation}
\begin{figure}[tbp]
  \begin{center}
    \includegraphics[width=6.5cm]{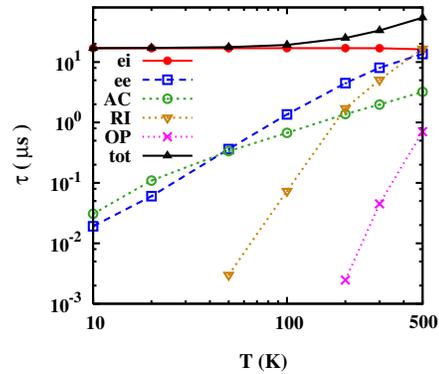}
  \end{center}
  \caption{(Color online) Total SRT ($\blacktriangle$) and the
    contributions from the electron-impurity ($\bullet$),
    electron-electron Coulomb ($\square$), electron--AC-phonon 
    ($\circ$), electron--RI-phonon ($\blacktriangledown$) as well as
    electron--OP-phonon ($\times$) scatterings as function of
    temperature $T$.}
  \label{fig_T}
\end{figure}

In Fig.~\ref{fig_T}, we plot the total SRT and the contributions from
each individual scattering as function of temperature $T$. It is seen
that the SRT changes little with temperature when $T$ varies from $5$~K
to $100$~K, and increases with increasing temperature when $T>100$~K. 
The underlying physics is as follows.
As said above, the inhomogeneous broadening does not change
with temperature and electron density, therefore the temperature
dependence of the SRT is determined by the momentum scattering:
stronger momentum scattering leads to longer SRT in the strong
scattering limit.\cite{Wu_rev} 
The electron-impurity scattering, which dominates the momentum
scattering at low temperature, depends weakly on temperature. 
Thus the SRT varies very mildly with $T$.
However, the electron-phonon and electron-electron Coulomb scatterings
both increase with temperature,\cite{Zhou_PRB_07,Vignale,Ivchenko} 
and become comparable to the impurity scattering at high temperature.
This enhances the momentum scattering and hence increases the SRT. 
It is noted that the electron-electron Coulomb scattering, 
which is absent in the previous investigations on spin relaxation in
graphene,\cite{Fabian_SR,Hernando_07,Hernando_09}
plays an important role in spin relaxation at high temperature.
We also show that the electron--OP-phonon scattering is always
negligible in the parameter regime of our investigation, which is
consistent with the claim in the previous
literature.\cite{Sarma_AC,Chen_Nnano}  

\begin{figure}[tbp]
  \begin{center}
    \includegraphics[width=6.5cm]{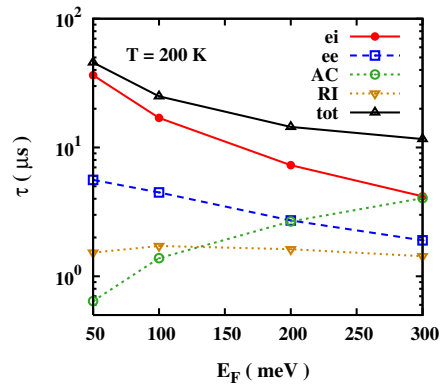}
  \end{center}
  \caption{(Color online) Total SRT ($\blacktriangle$) and the
    contributions from the electron-impurity ($\bullet$),
    electron-electron Coulomb ($\square$), electron--AC-phonon 
    ($\circ$) as well as electron--RI-phonon ($\blacktriangledown$)
    scatterings as function of Fermi energy $E_{\rm F}$.
$T=200$~K.}
  \label{fig_Ne}
\end{figure}

Then we turn to the electron-density dependence. In
Fig.~\ref{fig_Ne}, the total SRT and the contributions from various
scatterings are plotted against the Fermi energy $E_{\rm F}$ at 
$T=200$~K.\cite{noRI} It is seen that the SRT decreases first rapidly
and then slowly with increasing $E_{\rm F}$. 
To understand the underlying physics, we first discuss the
electron-density dependence of each individual scattering. 
The electron-impurity scattering decreases with $E_{\rm F}$ due to 
the decrease of the cross section.\cite{Guinea_08}
The electron-electron Coulomb scattering also decreases with $E_{\rm F}$ 
in the degenerate regime due to the increase of the Pauli
blocking.\cite{Jiang_bulk_09}
The electron--AC-phonon scattering increases with increasing $E_{\rm F}$
since the matrix element ($\sim q$) and the density of states
($\sim k$) both increase with $E_{\rm F}$.\cite{Sarma_AC} 
The electron--RI-phonon scattering varies slowly with $E_{\rm F}$ due to the
competition of the decrease in the matrix element and the increase in the
density of states.\cite{Fratini_RI}
Under the joint effects of these factors, 
the behaviour in $\tau-E_{\rm F}$ curve is understood: 
the SRT first decreases rapidly with $E_{\rm F}$ due to the decrease of the
electron-impurity scattering, and then decreases slowly since the increase of
the electron--AC-phonon scattering partially compensates the
decrease of the impurity scattering.

\begin{figure}[tbp]
  \begin{center}
    \includegraphics[width=6.5cm]{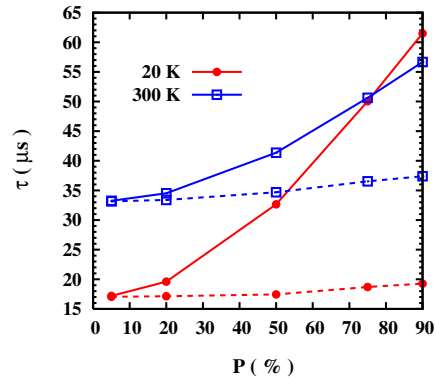}
  \end{center}
  \caption{(Color online) SRT as function of initial spin polarization $P$
    with (solid curves) and without (dashed curves) the
Coulomb HF term at $T=20$~K ($\bullet$) and 300~K ($\square$). }
  \label{fig_Hp}
\end{figure}

The initial spin polarization dependence of the SRT is also investigated. In
Fig.~\ref{fig_Hp}, we plot the SRT versus initial spin polarization $P$
at $T=20$~K and $200$~K. It is seen that the SRT
increases rapidly with the increase of the initial spin polarization. 
By comparing the calculation with and without the Coulomb HF term, 
one can see that the increase of the SRT originates from the
Coulomb  HF term. 
The underlying physics is similar to the previous studies
in semiconductors:\cite{highP} the Coulomb HF term serves as an
effective magnetic field along the $z$-axis, which is described by 
\begin{equation}
  B_{\rm HF}({\bf k})=\sum_{\mu{\bf k}^\prime}\left.V^{\mu 11}_{{\bf k},{\bf k}^\prime}
  I_{{\bf k}{\bf k}^\prime}(f_{\mu{\bf k}^\prime\frac{1}{2}}
  - f_{\mu{\bf k}^\prime-\frac{1}{2}})\right/(g \mu_B).
  \label{B_HF}
\end{equation}
This effective magnetic field blocks the spin precession and
slows down the spin relaxation. It is also shown that the SRT increases
slower with $P$ when temperature increases. This is because
at high temperature the electrons are distributed to a wider range in 
${\bf k}$ space, thus the effective magnetic field becomes smaller
[see Eq.~(\ref{B_HF})] and the effect of the HF term is weakened.\cite{highP} 
It is noted that in graphene there is a considerable increase of the
SRT with the initial spin polarization, even at room temperature.
In contrast, in semiconductors the electron system is in the nondegenerate
regime at room temperature (as the Fermi energy in semiconductor is
only tens of meV), and thus the effect of the HF term becomes
insignificant.\cite{highP}
This means that the HF effective field is more pronounced
in graphene compared with semiconductors.
Since the effects of the HF term have been probed experimentally in
semiconductors recently,\cite{wu-exp-hP,tobias,Zheng_exp}
we expect they can be observed easily in graphene. 

\begin{figure}[tbp]
  \begin{center}
    \includegraphics[width=6.5cm]{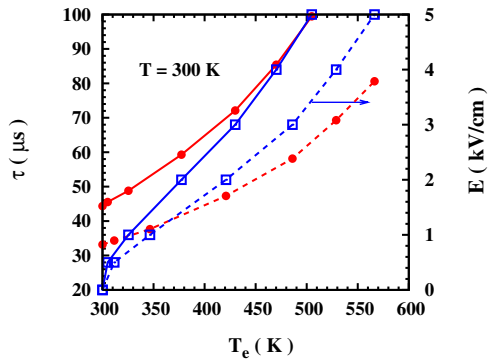}
  \end{center}
  \caption{(Color online) SRTs ($\bullet$) and in-plane electric field
    $E_{\|}$ ($\square$) as function of hot-electron temperature $T_e$ with (solid
    curves) and without (dashed curves) the electron-electron
Coulomb scattering at lattice temperature $T=300$~K and 
applied magnetic field in the Voigt configuration  $B=2$~T.}
  \label{fig_hot_e}
\end{figure}

We also study the effect of the in-plane electric field on spin
relaxation. In Fig.~\ref{fig_hot_e}, the SRT and the in-plane electric
field $E_{\|}$ are plotted against the hot-electron temperature $T_e$
for lattice temperature $T=300$~K and applied magnetic field $B=2$~T.
The hot-electron temperature is obtained by averaging the inverse of the 
slopes of $g_{\mu{\bf k}s}\equiv\log({1/f_{\mu{\bf k}s}^{\rm st}}-1)$ 
with ${\bf k}$ varying along different directions. 
Here $f_{\mu{\bf k}s}^{\rm st}$ represents the hot-electron
distribution function in the steady state, whose expression will be
discussed in the next subsection.  
From Fig.~\ref{fig_hot_e}, it is seen that the SRT increases monotonically with
the electric field, which is very different from the complicated behavior in 
semiconductor quantum wells.\cite{hot-e,Zhou_PRB_07}
In those systems, the high electric field induces two effects: (i) the
drift of the electron distribution which enhances the inhomogeneous broadening
as more electrons are distributed at larger $k$ and the SOC increases with $k$;
(ii) the hot-electron effect which enhances the momentum scattering.
The former tends to decrease the SRT while the latter tends to
increase. Therefore the electric field dependence of the
SRT can be nonmonotonic.\cite{hot-e,Zhou_PRB_07}
However, the spin-orbit field in graphene is independent on magnitude
of ${\bf k}$. Thus Effect (i) on spin relaxation
is marginal, and the electric field dependence of the SRT is mainly from
the hot-electron effect. Consequently the SRT increases 
{\it monotonically} with $E_{\|}$. In addition, by
comparing the calculation with and without the electron-electron
Coulomb scattering at the same hot-electron temperature, one finds
that the contribution to the SRT from the electron-electron scattering
increases with increasing hot-electron temperature, which is
consistent with the lattice temperature dependence discussed above.

\subsection{Steady-state hot-electron distribution function}
\begin{figure}[tbp]
  \begin{center}
    \includegraphics[width=6.2cm]{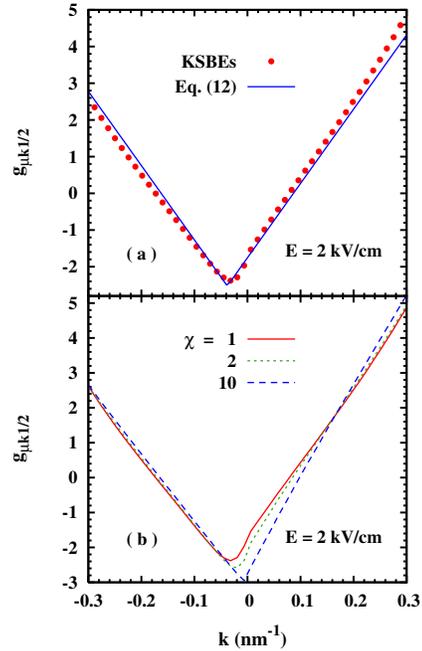}
  \end{center}
  \caption{ (Color online) (a) $g_{\mu{\bf k}\frac{1}{2}}\equiv
    \log({1/f^{\rm st}_{\mu{\bf k}\frac{1}{2}}}-1)$ from the KSBE computation
    (red dots) and from Eq.~(\ref{dis_weak}) (blue solid
    curve) against ${\bf k}$ varying along the direction of the
    in-plane electric field at $E_{\|}=2$~kV$/$cm and  $T=300$~K. 
    (b) $g_{\mu{\bf k}\frac{1}{2}}$ {\it vs.} ${\bf k}$ with the
    scaling coefficients $\chi=1$ (red solid curve), 2 (green
    dotted curve) and 10 (blue dashed curve).}
  \label{fig_dis}
\end{figure}

Previous investigations have shown that in the Boltzmann limit, the
strong electron-electron Coulomb scattering can establish the
steady-state hot-electron distribution function\cite{distribution}  
\begin{equation}
  \tilde f^{\rm st}_{\mu{\bf k}s}=\left\{\exp\left[
        (\varepsilon_{{\mu}{\bf k}} -{\bf u}\cdot{\bf k}-\mu_s)/(k_BT_e)
    \right] + 1  \right\}^{-1},
  \label{dis_strong}
\end{equation}
where $\mu_s$ stands for the chemical potential of electrons with spin
$s$ and ${\bf u}$ is the drift velocity.
Recently Bistritzer and MacDonald applied this distribution function to
study the charge transport in graphene.\cite{Bistritzer} 
However, whether the Coulomb scattering in graphene is strong enough to
justify the validity of this distribution function remains unchecked.
Since we can obtain the distribution function with the genuine
Coulomb scattering explicitly computed, we check the validity of
Eq.~(\ref{dis_strong}) here. 
It is noted, if  Eq.~(\ref{dis_strong}) is valid, 
 $g_{\mu{\bf k}s}\equiv \log({1/\tilde f_{\mu{\bf k}s}^{\rm st}}-1)=
(v_{\rm F}|{\bf k}|-{\bf u}\cdot{\bf k}-\mu_s)/k_BT_e$ 
has a minimum around $k=0$ and different slopes along different $k$-directions.
In Fig.~\ref{fig_dis}(a), $g_{\mu{\bf k}s}$ is plotted against
${\bf k}$ varying along the 
direction of the electric field at $E_{\|}=2$~kV$/$cm and $T=300$~K. 
One immediately finds that the hot-electron distribution function
from our calculation is very different from Eq.~(\ref{dis_strong}):
the minimum of $g_{\mu{\bf k}s}$ is away from the point of $k=0$ and
the slopes are close to each other when ${\bf k}$ varying along opposite
directions. Based on the above property, we propose the approximate
expression of the computed hot-electron distribution function as:
\begin{equation}
f^{\rm st}_{\mu{\bf k}s}=\left\{\exp\left[
        (\varepsilon_{{\mu}{\bf k}-{\bf u}}-\mu_s)/k_BT_e
    \right] + 1  \right\}^{-1}.
  \label{dis_weak}
\end{equation}
Correspondingly, 
$g_{\mu{\bf k}s}=(v_{\rm F}|{\bf k}-{\bf u}|-\mu_s)/k_BT_e$.
To examine our assumption, we also plot $g_{\mu{\bf k}s}$ from
Eq.~(\ref{dis_weak}) in Fig.~\ref{fig_dis}(a) and find that the
computed hot-electron distribution function is in reasonable agreement with
Eq.~(\ref{dis_weak}).
In fact, for systems with parabolic energy dispersion, e.g., semiconductors
in our previous investigations,\cite{hot-e,Jiang_bulk_09} 
Eqs.~(\ref{dis_strong}) and (\ref{dis_weak}) are equivalent. However,
for graphene with linear dispersion, these two distribution functions
are quite distinct. 
It is noted that Eq.~(\ref{dis_weak}) is just used to estimate the
hot-electron temperature. The SRT and hot-electron
distribution in this investigation are {\it explicitly} computed from the KSBEs.

In order to reveal the effect of Coulomb scattering to 
the steady-state hot-electron distribution,
we introduce a dimensionless scaling coefficient $\chi$
in front of the electron-electron Coulomb scattering, with 
 $\chi=1$ corresponding to the genuine case. We plot $g_{\mu{\bf k}s}$
against ${\bf k}$ along the direction of the electric field
with different scaling coefficients $\chi$ in Fig.~\ref{fig_dis}(b).
It is seen that with the increase of
$\chi$, i.e., the electron-electron Coulomb scattering, the
distribution function gets closer  to Eq.~(\ref{dis_strong}). 
This indicates that in graphene the electron-electron Coulomb scattering
is not strong enough to establish the hot-electron distribution 
Eq.~(\ref{dis_strong}).

\section{Conclusion  and Discussion}
In conclusion, we have investigated the spin relaxation in graphene from the
microscopic KBSE approach, where all the relevant scatterings,
especially the electron-electron Coulomb scattering, are explicitly included.
We show that the SRT remains almost unchanged with increasing $T$ at
low temperature because the electron-impurity scattering, which dominates the
momentum scattering, varies little with temperature. Nevertheless, the SRT
increases with $T$ at high temperature because the electron-electron and
electron-phonon scatterings  become comparable to the
electron-impurity scattering and both scatterings increase with increasing $T$.
We also show that the electron-electron
Coulomb scattering plays an important role in spin 
relaxation at high temperature.
It is also seen that the SRT first decreases 
rapidly with the increase of $E_{\rm F}$
due to the decrease of the electron-impurity scattering,
and then decreases mildly
with $E_{\rm F}$ since the increase of the electron--AC-phonon scattering 
partially counteracts  the decrease of the electron-impurity scattering.
We also predict a pronounced increase of the SRT at high spin polarization
in the {\it whole} temperature regime of our investigation. The
underlying physics is that the Coulomb HF term serves as an effective
longitudinal magnetic field which blocks the spin precession and suppresses the
spin relaxation. The effect of the in-plane electric field on
spin relaxation is also investigated. It is shown that the SRT
increases with the in-plane electric field due to the hot-electron effect. 
Moreover, we show that the electron-electron Coulomb scattering
in graphene  is not strong enough to establish the usual steady-state
hot-electron distribution in the literature and an  
approximate one is suggested based on our computation. 

Now we address the effect of ripples on the
spin relaxation.  For graphene samples with an
undulating surface, i.e. ripples,\cite{Morozov_curv} an
additional Rashba-type SOC, whose expression is the same 
as the electric field induced SOC,\cite{Hernando_06} appears. The SOC
coefficient due to the curvature effect is estimated to be
$\alpha_{\rm curv}{\sim} 0.02$~meV (0.2~K),\cite{Hernando_06} which is
about two orders of magnitude larger than 
$\alpha_{\rm R}=1.5\times10^{-4}$~meV used in our calculations.
For the well-known relation $1/\tau\propto \alpha_{\rm R}^2$ for the DP
mechanism,\cite{Dyakonov} one 
expects that the SRT is shortened  by four orders of magnitude.
Nevertheless, the additional SOC does not change the dependences 
of the SRT on the temperature and sample parameters,
as well as the importance of the electron-electron Coulomb scattering.

Finally, it is noted that even after considering the SOC enhanced by
the curvature effect, the SRTs from our calculation are 
still two orders of magnitude larger than those in recent spin transport 
experimental measurements.\cite{Tombros_07,Tombros_08,Popinciuc,Jozsa_09} 
We  stress that the range of impurity (adatom) density of
the graphene samples we discuss is different from that in recent experiments,
and therefore the dominant spin relaxation mechanism is also quite different.
As mentioned above, the mechanism most likely limiting the SRTs in the recent
experiments is the local spin-orbit field induced by the randomly distributed
adatoms. The experimental and theoretical works\cite{Varykhalov,Castro_imp}  
showed that the SOC strength from the adatoms can reach $10$~meV,
which is about three orders of magnitude larger than the strength used
in our calculations. The study on the effect of the adatoms on
spin relaxation is beyond the scope of this investigation.
It is further noted that the spin-related 
experiment in clean graphene is still missing, 
we expect that the effects presented in this manuscript can be
confirmed by the future experiments in relatively cleaner graphene
samples. A possible method to obtain the graphene sample with higher
mobility is the epitaxial growth of graphene on SiC
substrate.\cite{Emtsev,Virojanadara,Deretzis} 
Nevertheless in this system, some other factors must be
taken into account, e.g., the scattering arising
from the interfacial states.\cite{Deretzis} The study 
in this system can be the future
extension of this investigation.

\begin{acknowledgments}
This work was supported by the National Natural Science Foundation of
China under Grant No.\ 10725417, the National Basic
Research Program of China under Grant No.\ 2006CB922005 and the
Knowledge Innovation Project of Chinese Academy of Sciences.
The authors acknowledge valuable discussions with J. Fabian.
\end{acknowledgments}

\appendix*
\section{Expression of the interaction Hamiltonian}
The electron-impurity and electron-electron Coulomb interaction 
Hamiltonian can be written as 
\begin{widetext}
\begin{eqnarray}
  H_{ei}&=&\sum_{j{\mu}{\nu} \atop {\bf k}{\bf q}s} 
  U^{\mu\nu}_{{\bf k},{\bf k}-{\bf q}}
  T^{{\mu}\nu\nu}_{{\bf k}{\bf k}-{\bf q}}
  e^{-i{\bf q}\cdot {\bf R}_j} {c^{{\mu}\nu}_{{\bf k}s}}^{\dagger}
  c^{{\mu}\nu}_{{\bf k}-{\bf q}s}, \\ 
  H_{ee}&=&\frac{1}{2}\sum_{{\mu}{\mu}^\prime \nu\nu^\prime
    \atop {\bf k}{\bf k}^\prime{\bf q}ss^\prime} 
  V^{\mu\nu\nu}_{{\bf k},{\bf k}-{\bf q}} 
  T^{{\mu}\nu\nu}_{{\bf k}{\bf k}-{\bf q}}
  T^{{\mu}^\prime\nu^\prime\nu^\prime}_{{\bf k}^\prime{\bf k}^\prime+q}
  {c^{{\mu}\nu}_{{\bf k}s}}^{\dagger}
  {c^{{\mu}^\prime\nu^\prime}_{{\bf k}^\prime s^\prime}}^{\dagger}
  c^{{\mu}^\prime\nu^\prime}_{{\bf k}^\prime+qs^\prime} 
  c^{{\mu}\nu}_{{\bf k}-{\bf q}s} \nonumber\\
&&\mbox{}\hspace{3cm}  + \frac{1}{2}\sum_{{\mu}{\mu}^\prime, \nu\neq\nu^\prime
    \atop {\bf k}{\bf k}^\prime{\bf q}ss^\prime} 
  V^{\mu\nu\nu^\prime}_{{\bf k},{\bf k}-{\bf q}} 
  T^{{\mu}\nu\nu^\prime}_{{\bf k}{\bf k}-{\bf q}}
  T^{{\mu}^\prime\nu^\prime\nu}_{{\bf k}^\prime{\bf k}^\prime+q}
  {c^{{\mu}\nu}_{{\bf k}s}}^{\dagger}
  {c^{{\mu}^\prime\nu^\prime}_{{\bf k}^\prime s^\prime}}^{\dagger}
   c^{{\mu}^\prime\nu}_{{\bf k}^\prime+qs^\prime} 
  c^{{\mu}\nu^\prime}_{{\bf k}-{\bf q}s}.
\end{eqnarray}
\end{widetext}
In these equations, ${\bf R}_j$ stands for the position of $j$th
impurity; $T^{{\mu}\nu\nu^\prime}_{{\bf k}{\bf k}-{\bf q}}=
{\psi_{{\bf k}}^{\mu\nu}}^\dagger\psi_{{\bf k}-{\bf q}}^{\mu\nu^\prime}$;
$U^{\mu\nu}_{{\bf k},{\bf k}-{\bf q}}=Z_i
V^{\mu\nu\nu}_{{\bf k},{\bf k}-{\bf q}}e^{-qd}$ is the
electron-impurity interaction matrix element. Here
$Z_i=1$ is the charge number of the impurity; the effective
distance $d$ of the impurity layer to the graphene sheet is chosen to
be $0.4$~nm.\cite{Fabian_SR,Adam_07,Hwang_PRL,Adam_08,Fratini_RI}
$V^{\mu\nu\nu^\prime}_{{\bf k},{\bf k}-{\bf q}}$ denotes the screened Coulomb
potential where the screening is calculated under the random phase
approximation,\cite{Manhan,Jonson,Zhou_PRB_07,Haug_1998,Adam_07,
Hwang_PRL,Adam_08,Sarma_Coulomb,Sarma_RPA,Wunsch,Ramezanali,WangXF}
\begin{equation}
  V^{\mu\nu\nu^\prime}_{{\bf k},{\bf k}-{\bf q}} =\frac{V_{{\bf q}}^{(0)}}{
  {1-V_{{\bf q}}^{(0)}\Pi({\bf q},\epsilon_{{\mu}\nu{\bf k}}-
    \epsilon_{{\mu}\nu^\prime{\bf k}-{\bf q}})}},
\end{equation}
where $V_{{\bf q}}^{(0)}={{2\pi v_{\rm F} r_s}}/{q}$ is the two-dimensional bare Coulomb
potential with $r_s=0.8$.\cite{Adam_07,Hwang_PRL,Adam_08}
As pointed out in Refs.~\onlinecite{Adam_07} and \onlinecite{Hwang_PRL}, 
such small $r_s$ ensures the validity of the random phase approximation. 
$\Pi(\mathbf{q},\omega)$ is given by\cite{Wunsch,Sarma_RPA,WangXF,Ramezanali}
\begin{align}
  \Pi(\mathbf{q},\omega)=\sum_{\mu\nu\nu^{\prime}{\bf  k}s}
  |T^{{\mu}\nu\nu^\prime}_{{\bf k}{\bf k}-{\bf q}}|^2
  \frac{f^{\mu\nu}_{{\bf k}s}-f^{\mu\nu^{\prime}}_{{\bf k}-{\bf q}s}}
  {\epsilon_{\mu\nu{\bf k}}-\epsilon_{\mu\nu^\prime{\bf k}-{\bf q}}
    +\omega +i0^+}.
  \label{eq:P1}
\end{align}
It is noted that the interband contribution in screening cannot be
neglected even in the $n$-doped case.

The electron-phonon interaction Hamiltonian takes the form
(only the terms relevant to the electron band presented)
\begin{equation}
  H_{ep}=\sum_{{\mu}{\mu}^\prime,\nu=1 \atop {\bf k}{\bf q}s} 
  M^{\lambda}_{\mu{\bf k},\mu^\prime{\bf k}-{\bf q}}
  (a_{{\lambda},{\bf q}}+ a_{{\lambda},-{{\bf q}}}^\dagger) 
  {c^{{\mu}\nu}_{{\bf k}s}}^{\dagger} 
  c^{{\mu}^\prime\nu}_{{\bf k}-{\bf q}s}.
\end{equation}
Here $a_{{\lambda},{\bf q}}$ (${a_{{\lambda},{\bf q}}}^\dagger$) is the
annihilation (creation) operator and
$M^{\rm \lambda}_{\mu{\bf k},\mu^\prime{\bf k}-{\bf q}}$ stands for the matrix
element of the electron-phonon interaction 
with $\lambda$ being the phonon branch index.
For the electron--AC-phonon scattering, the 
phonon energy spectrum $\omega^{\rm AC}_{\bf q}=v_{\rm ph}q$
and $|M^{\rm AC}_{\mu{\bf k},\mu^\prime{\bf k}-{\bf q}}|^2=
\frac{D^2 q}{2\rho_{m}v_{\rm ph}}
I_{{\bf k}{\bf k}-{\bf q}}\delta_{\mu\mu^\prime}$, where
$v_{\rm ph}=2\times 10^6$~cm/s is acoustic phonon velocity and
$\rho_{m}=7.6\times 10^{-8}$~g/cm$^2$ denotes graphene mass
density.\cite{Sarma_AC,Chen_Nnano}
The value of deformation potential $D$ is still in debate. The range of
$D$ obtained via various theoretical and experimental methods is from $4.5$ to 
$30$~eV.\cite{Chen_Nnano,Borysenko,Bolotin,Chen_AC,Hong_AC,Stauber,Pietronero,Sugihara}
Here we choose a moderate value $D=19$~eV.\cite{Sarma_AC,Chen_Nnano,Sugihara}
For the electron--RI-phonon scattering,
$|M^{\rm RI}_{\mu{\bf k},\mu^\prime{\bf k}-{\bf q}}|^2=
g\frac{v_{\rm F}^2}{a}\frac{e^{-2qd}}{q+q_s}
I_{{\bf k}{\bf k}-{\bf q}}\delta_{\mu\mu^\prime}$ where
$q_s=4r_sk_{\rm F}$ is the Thomas-Fermi screening length,\cite{Wunsch}
$a=1.42$~\AA\ is the C-C bond distance.
For SiO$_2$ substrate, the energy of the remote phonon modes are
$\omega^{\rm RI}_{1}=59$~meV and $\omega^{\rm RI}_{2}=155$~meV; the 
corresponding dimensionless coupling parameters are $g_1=5.4\times 10^{-3}$
and $g_2=3.5\times 10^{-2}$, respectively.\cite{Fratini_RI}
The matrix element of the electron--OP-phonon scattering is described by
$|M^{\rm OP}_{\mu{\bf k},\mu^\prime{\bf k}-{\bf q}}|^2=\frac{A_{\lambda}}
{2\rho_{m}\omega_{\lambda}}$.
For the longitudinal and transverse optical phonons near $\Gamma$ point,
which cause the intravelley scattering,
$A^{\rm LO}_{\Gamma}=\langle D_{\Gamma}^2\rangle[1-\cos(\theta_{\bf k}
+\theta_{\bf k-q}-2\theta_{\bf q})]\delta_{\mu\mu^\prime}$,
$A^{\rm TO}_{\Gamma}=\langle D_{\Gamma}^2\rangle[1+\cos(\theta_{\bf k}
+\theta_{\bf k-q}-2\theta_{\bf q})]\delta_{\mu\mu^\prime}$ with $\langle
D_{\Gamma}^2\rangle=45.60$~eV$^2$/\AA$^2$ and $\omega_{\Gamma}=196.0$~meV;
whereas for the transverse optical phonon
near $K(K^\prime)$ point which causes the intervalley scattering,
$A^{\rm TO}_{K}=\langle D_{K}^2\rangle
[1-\cos(\theta_{\bf k}-\theta_{\bf k-q})]\delta_{\mu,-\mu^\prime}$ with
$\langle D_{K}^2\rangle=92.05$~eV$^2$/\AA$^2$ and
$\omega_{K}=161.2$~meV.\cite{Piscanec_OP,Lazzeri_OP}


\begin{thebibliography}{0}
\bibitem{noRI} The contribution on spin relaxation from the electron--RI-phonon
  phonon is not shown here since it is much smaller than those from other
  scatterings at $T=200$~K.
\end{thebibliography}
\end{document}